\documentclass[a4paper,11pt]{article}
\pdfoutput=1 

\usepackage[margin=1.0in]{geometry}

\usepackage[utf8]{inputenc}
\usepackage{graphics}
\usepackage{graphicx}
\usepackage{url}
\usepackage{caption} 
\usepackage{amsmath}
\usepackage[merge,numbers,compress]{natbib}
\usepackage{hyperref}
\usepackage{placeins}
\usepackage{morefloats}
\usepackage{slashed}
\usepackage[colorinlistoftodos, shadow]{todonotes}
\usepackage{titlesec}
\usepackage{gensymb}
\usepackage{textcomp}
\usepackage{amssymb}
\usepackage{verbatim}
\usepackage{xspace}

\allowdisplaybreaks

\allowdisplaybreaks[1]

\def\mathswitch#1{\relax\ifmmode#1\else$#1$\fi}
\def\mathswitchr#1{\relax\ifmmode{\mathrm{#1}}\else$\mathrm{#1}$\fi}
\def\mathswitchit#1{\relax\ifmmode{#1}\else$#1$\fi}

\newcommand{\Pp}{\mathswitchr p}

\newcommand{\Pep}{\mathswitchr {e^+}}
\newcommand{\Pem}{\mathswitchr {e^-}}

\newcommand{\Pmum}{\mathswitchr {\mu^-}}

\newcommand{\PW}{\mathswitchr W}
\newcommand{\PZ}{\mathswitchr Z}

\newcommand{\Pg}{\mathswitchr g}

\newcommand{\Pne}{\mathswitch \nu_{\mathrm{e}}}

\newcommand{\sw}{\mathswitch {s_{\mathrm{w}}}}
\newcommand{\cw}{\mathswitch {c_{\mathrm{w}}}}

\newcommand{\cgw}{ g_{\rm w}}

\def\bfi{\begin{figure}}
\def\efi{\end{figure}}

\newcommand{\im}{{\rm  i}}

\def\draftdate{\relax}
\def\mda{\relax}
\def\mua{\relax}
\def\mla{\relax}
\def\draft{
  \def\thtystars{******************************}
  \def\sixtystars{\thtystars\thtystars}
  \typeout{}
  \typeout{\sixtystars**}
  \typeout{* Draft mode!
    For final version remove \protect\draft\space in source file *}
  \typeout{\sixtystars**}
  \typeout{}
  \def\draftdate{\today}
  \def\mua{\marginpar[\boldmath\hfil$\uparrow$]%
    {\boldmath$\uparrow$\hfil}%
    \typeout{marginpar: $\uparrow$}\ignorespaces}
  \def\mda{\marginpar[\boldmath\hfil$\downarrow$]%
    {\boldmath$\downarrow$\hfil}%
    \typeout{marginpar: $\downarrow$}\ignorespaces}
  \def\mla{\marginpar[\boldmath\hfil$\rightarrow$]%
    {\boldmath$\leftarrow $\hfil}%
    \typeout{marginpar: $\leftrightarrow$}\ignorespaces}
  \def\Mua{\marginpar[\boldmath\hfil$\Uparrow$]%
    {\boldmath$\Uparrow$\hfil}%
    \typeout{marginpar: $\uparrow$}\ignorespaces}
  \def\Mda{\marginpar[\boldmath\hfil$\Downarrow$]%
    {\boldmath$\Downarrow$\hfil}%
    \typeout{marginpar: $\downarrow$}\ignorespaces}
  \def\Mla{\marginpar[\boldmath\hfil$\Rightarrow$]%
    {\boldmath$\Leftarrow $\hfil}%
    \typeout{marginpar: $\leftrightarrow$}\ignorespaces}
  \overfullrule 5pt
  \oddsidemargin -15mm
  \marginparwidth 29mm
}

\numberwithin{equation}{section}

\begin{document} 

\thispagestyle{empty}
\def\thefootnote{\fnsymbol{footnote}}
\setcounter{footnote}{1}
\null

\vfill
\begin{center}
  {\Large {\boldmath\bf {Effects of anomalous triple-gauge-boson interactions in diboson 
  production with {\sc RECOLA2}
        \footnote{Talk presented at Loops and Legs in Quantum Field Theory (LL2018), 29 April 2018 - 04 May 2018, St. Goar, Germany.
          M.~C. would like to thank the organizers for their kind invitation.}}
      \par} \vskip 2.5em
    {\large
      {\sc Mauro Chiesa$^{1}$, Ansgar Denner$^{1}$, Jean-Nicolas Lang$^{2}$
      }\\[2ex]
      {\normalsize \it
        $^1$Julius-Maximilians-Universit\"at W\"urzburg,
        Institut f\"ur Theoretische Physik und Astrophysik, \\
        Emil-Hilb-Weg 22, D-97074 W\"urzburg, Germany
      }\\[2ex]
      {\normalsize \it
        $^2$Universit\"at Z\"urich, Physik-Institut, CH-8057 Z\"urich,
        Switzerland}
    }
  }
  \par \vskip 1em
\end{center}\par
\vskip .0cm \vfill {\bf Abstract:}
\par
Diboson production processes are of great importance in high-energy physics
  and a precise theoretical knowledge of $VV^{(')}$ production is mandatory not only in view of precision
  tests of the Standard Model but also in the one of new physics searches. We present results of Ref.~\cite{Chiesa:2018lcs}, where we
  computed the NLO QCD and NLO electroweak
  corrections to diboson production processes at the LHC including the effect of the anomalous triple-gauge-boson
  interactions at NLO QCD accuracy. The anomalous triple-gauge-boson interactions are parametrized in terms of
  higher-dimensional operators in the effective field theory framework. Our calculation is the first application
  of {\sc Recola2} to effective field theory models.
\par
\vskip 1cm
\noindent
\par
\null \setcounter{page}{0} \clearpage
\def\thefootnote{\arabic{footnote}} \setcounter{footnote}{0}


\section{Introduction}
\label{sec:Intro}

Diboson production is of great importance at the LHC. On the one hand, it can be used as a precision test of the Standard Model (SM)
and, on the other hand, it is an important source of background both in the context of SM measurements and in the one of New Physics (NP) searches.
Moreover, diboson production is sensitive to the gauge-boson self interaction and can be used to set limits on the anomalous triple-gauge-boson
interactions.

From the above considerations, it follows that a precise theoretical understanding of this process is  mandatory. The current status of
the theoretical predictions for diboson production can be summarized as follows. On the QCD side, the two-loop QCD corrections to $\Pp \Pp \to V V' (\to 4l)$
have been computed, resummed predictions at NNLL+NNLO accuracy and matched predictions at NLO+PS (Parton Shower) accuracy are also available. More recently, an NNLO
calculation matched with QCD PS was presented in Ref.~\cite{Re:2018vac}. On the electroweak (EW) side, the one-loop EW corrections have been computed for all the
four-lepton production processes. We refer to Ref.~\cite{Chiesa:2018lcs} for a more detailed bibliography.

\section{EFT framework for triple-gauge-boson interaction}
\label{sec:EFT}

In the literature, the anomalous triple-gauge-boson interaction is usually described in terms of anomalous couplings. More recently
have been published 
results in the SM effective field theory (SMEFT) framework.
In the SMEFT, the SM Lagrangian is generalized by adding non-renormalizable gauge-invariant operators
with canonical dimension $D>4$:
\begin{equation}
  {\cal L}^{\rm eff.}={\cal L}^{\rm SM} + \sum_i \frac{c^i_6}{\Lambda^2}{\cal O}^i_6
  + \sum_i \frac{c^i_8}{\Lambda^4}{\cal O}^i_8 + \cdots .
  \label{eq:eftgen}
\end{equation}
In Eq.~(\ref{eq:eftgen}) the operators ${ \cal O }^i_D$ represent the
effect of new physics with a mass scale $\Lambda$ much larger than the
electroweak scale and are multiplied by the corresponding Wilson
coefficients $c^i_D$.

We describe the anomalous $WW\gamma$ and $WWZ$ interaction in terms of the following dimension-6 (Dim-6) operators~\cite{Degrande:2012wf}:
\allowdisplaybreaks
\begin{equation}
  \begin{aligned}
    {\cal O}_{WWW}&= - \frac{\cgw^3}{4} \epsilon_{ijk} W_{\mu\nu}^i W^{\nu\rho\;j}W_{\rho}^{~\mu\;k},\\
    {\cal O}_W&    = - \im \cgw (D_\mu\Phi)^\dagger \frac{\tau_k}{2} W^{\mu\nu\;k}(D_\nu\Phi), \\
    {\cal O}_B&    = + \im \frac{g_1}{2} (D_\mu\Phi)^\dagger B^{\mu\nu}(D_\nu\Phi), \\
    {\cal O}_{\widetilde{W}WW}&=  + \frac{\cgw^3}{4} \epsilon_{ijk} {\widetilde{W}}_{\mu\nu}^i W^{\nu\rho\;j}W_{\rho}^{~\mu\;k},\\
    {\cal O}_{\widetilde{W}}&  =  + \im \cgw (D_\mu\Phi)^\dagger \frac{\tau_k}{2} {\widetilde{W}}^{\mu\nu\; k}(D_\nu\Phi) , 
  \end{aligned}
  \label{eq:tgcWWv}
\end{equation}
where $\cgw=e/\sw$, $g_1=e/\cw$, $\tau$ are the Pauli matrices and $\Phi$ stands for the Higgs doublet.

As no Dim-6 operator contributes to the neutral triple-gauge-boson interaction, we consider the following set
of Dim-8 operators for the $VVV'$ interaction ($V,V'=Z,\gamma$)~\cite{Degrande:2013kka}:
\begin{equation}
  \begin{aligned}
  \mathcal{O}_{BW}&= - \im \, \Phi^\dagger   B_{\mu\nu}
  \frac{\tau_i}{2} W^{\mu\rho\; i} \left\{D_\rho,D^\nu\right\} \Phi + \mathrm{h.c.}, \\
  \mathcal{O}_{WW}&= \im \, \Phi^\dagger \frac{\tau_i}{2} \frac{\tau_j}{2} W_{\mu\nu}^iW^{\mu\rho \;j} \left\{D_\rho,D^\nu\right\} \Phi + \mathrm{h.c.} , \\
  \mathcal{O}_{BB}&= \im \, \Phi^\dagger  B_{\mu\nu}B^{\mu\rho} \left\{D_\rho,D^\nu\right\} \Phi + \mathrm{h.c.} , \\
  \mathcal{O}_{\widetilde{B}W}&= - \im \, \Phi^\dagger  \widetilde{B}_{\mu\nu} \frac{\tau_i}{2} W^{\mu\rho \; i} \left\{D_\rho,D^\nu\right\} \Phi + \mathrm{h.c.},
  \end{aligned}
  \label{eq:ntgcop}
\end{equation}
where h.c. denotes the  hermitian conjugate, $D_{\mu}$ represents the
${\rm SU}(2)_{\rm  w} \times {\rm U}(1)_{Y}$ covariant derivative and
$\left\{D_\mu,D^\nu\right\}=D_{\mu}D^{\nu}+D^{\nu}D_{\mu}$.

As a consequence of Eq.~(\ref{eq:eftgen}), the cross sections and/or the differential distributions have the form
\begin{equation}
  \sigma = \sigma_{{\rm SM}^2} + \sigma_{{\rm SM}\times {\rm EFT6}} +
  \sigma_{{\rm EFT6}^2} + \sigma_{{\rm SM}\times {\rm EFT8}}  +
  \sigma_{{\rm EFT8}^2} + \dots ,  
  \label{eq:EFTobs}
\end{equation}
with
\begin{equation}
  \sigma_{{\rm SM}\times {\rm EFT6}} \propto \frac{c_6}{\Lambda^2} ,
  \quad  
\sigma_{{\rm EFT6}^2} \propto \frac{c_6^2}{\Lambda^4} \quad 
  \sigma_{{\rm SM}\times {\rm EFT8}} \propto \frac{c_8}{\Lambda^4}, \quad 
\sigma_{{\rm EFT8}^2} \propto \frac{c_8^2}{\Lambda^8}  .  
  \label{eq:EFTobs2}
\end{equation}
Since the $\sigma_{{\rm EFT6}^2}$ and $\sigma_{{\rm SM}\times {\rm EFT8}}$ terms are
of the same order in the $1/ \Lambda$ expansion, for a generic EFT model a consistent
$1/ \Lambda$ expansion should include only the $\sigma_{{\rm SM}\times {\rm EFT6}}$
contribution if the Dim-8 operators are neglected. However, there can be specific models
where the $\sigma_{{\rm SM}\times {\rm EFT8}}$ term is subleading with respect to the 
$\sigma_{{\rm EFT6}^2}$ one without violating the EFT expansion. For these reasons, in Sect.~\ref{sec:results} 
we present results both with and without the quadratic contribution in the Dim-6 operators.
Similar considerations hold for the Dim-8 operators in ZZ production.

The values of the Wilson coefficients that we used in our numerical simulations are consistent
with the current experimental limits from the LHC.

\section{Overview of the calculation}
\label{sec:calculation}

We computed $\PW\PW$~($\to \Pep \Pne \Pmum \bar{\nu}_{\mu}$), $\PW\PZ$~($\to \Pep \nu_{\rm e} \mu^+ \mu^-$),
and $\PZ\PZ$~($\to \Pep \Pem\mu^+ \mu^-$) production
at NLO QCD including the effect of the Dim-6 and Dim-8 operators as described in Sect.~\ref{sec:EFT}.
For the processes under consideration, also the NLO EW corrections have been calculated. For WW and ZZ
production, we included the effect of the loop-induced $\Pg\Pg \to \PW\PW$ and $\Pg\Pg \to \PZ\PZ$ processes;
though these processes are formally of higher order in $\alpha_{\rm S}$, they can give sizable
contributions because of  the gluon PDF.

Our calculation relies on the following tools: {\sc FEYNRULES}, {\sc REPT1L}, and {\sc RECOLA2}.
We used the {\sc MATHEMATICA} package {\sc FEYNRULES}~\cite{Christensen:2008py,Alloul:2013bka} to implement the SM Lagrangian and the relevant Dim-6 and
Dim-8 operators. We then employed {\sc FEYNRULES} to derive the corresponding Feynman rules which were written in a
UFO~\cite{Degrande:2011ua} model file.

The {\sc PYTHON} program {\sc REPT1L}~\cite{Denner:2017wsf} was used to convert the UFO model file to a {\sc RECOLA2} model file.
{\sc REPT1L} performs counterterm expansion, sets up and solves the renormalization conditions and computes 
the rational terms of type R2 for the model under consideration.

{\sc RECOLA2}~\cite{Denner:2017wsf,Denner:2017vms} is an enhanced version of {\sc RECOLA}~\cite{Actis:2016mpe}
employing the tensor-reduction library {\sc COLLIER}~\cite{Denner:2016kdg} and is used for the automated generation and the
numerical evaluation of the tree-level and one-loop amplitudes. 
While {\sc RECOLA} is designed for the SM, 
{\sc RECOLA2} relies on external {\sc RECOLA2} model files: these model files are user-provided and can be
generalizations of the SM. 

{\sc RECOLA2} was interfaced to a multichannel Monte Carlo integrator that performs all the other steps of the
calculation.

\section{Numerical results}
\label{sec:results}

For the process $\Pp \Pp \to \PW\PZ \to \Pep \nu_{\rm e} \mu^+ \mu^-$ we consider two event selections that mimic
the ATLAS and CMS analyses of Refs.~\cite{Aad:2016ett} and~\cite{Khachatryan:2016poo}, respectively. Some results at the cross-section level are collected in
Tab.~\ref{tab:wz-xsec}. It turns out that the NLO EW corrections are of order $-6\% / -5\%$, while the NLO QCD corrections are
positive and large ($+80\%/+90\%$): this is due to the fact that new channels appear at NLO QCD that have gluons in the initial state
and are enhanced by the gluon PDF.

\begin{table}
  \begin{center}
\renewcommand{\arraystretch}{1.4}
    \begin{tabular}{|l|l|l|l|l}
      \hline
       Setup & LO [fb] & NLO QCD [fb] & NLO EW [fb] \\
      \hline
      $\PW^+\PZ$ ATLAS & $18.875(1)^{+5.2\%}_{-6.4\%}$ & $34.253(6)^{+5.3\%}_{-4.3\%}$ & $17.748(8)^{+5.3\%}_{-6.5\%}$ \\
      \hline
      $\PW^+\PZ$   CMS & $14.307(1)^{+5.0\%}_{-6.2\%}$ & $26.357(6)^{+5.4\%}_{-4.3\%}$ & $13.600(4)^{+5.1\%}_{-6.3\%}$ \\
      \hline
    \end{tabular}
  \end{center} 
  \caption{Integrated cross section for $\PW\PZ$ production. In the first column $\PW^+\PZ$ 
    is a short-hand notation for the process $\Pp\Pp \to \Pep \nu_{\rm e} \mu^+ \mu^-$. 
    The numbers in parentheses correspond to the statistical
    error on the last digit. The uncertainties are estimated from the scale dependence.}
  \label{tab:wz-xsec}
\end{table}

Figure~\ref{fig:wzptmt} shows the differential distributions as a function of the Z boson $p_{\rm T}$
($p_{ {\rm T,}\mu^+ \mu^-}$) and of the WZ  transverse mass ($M_{\rm T}^{3l\nu}$). For these distributions,
the NLO EW corrections are negative and their size increases at high $p_{\rm T}$ and/or $M_{\rm T}$ up
to $-25\%$ and $-20\%$ in the tails of the distributions. The NLO QCD corrections are positive and very large,
in particular  for the $p_{ {\rm T,}\mu^+ \mu^-}$ distribution, where the NLO prediction is almost six times the LO one
for $p_{ {\rm T,}\mu^+ \mu^-} \simeq 600$~GeV.

\bfi
\begin{center}
  \begin{minipage}{0.40\textwidth}
    \includegraphics[width=\textwidth]{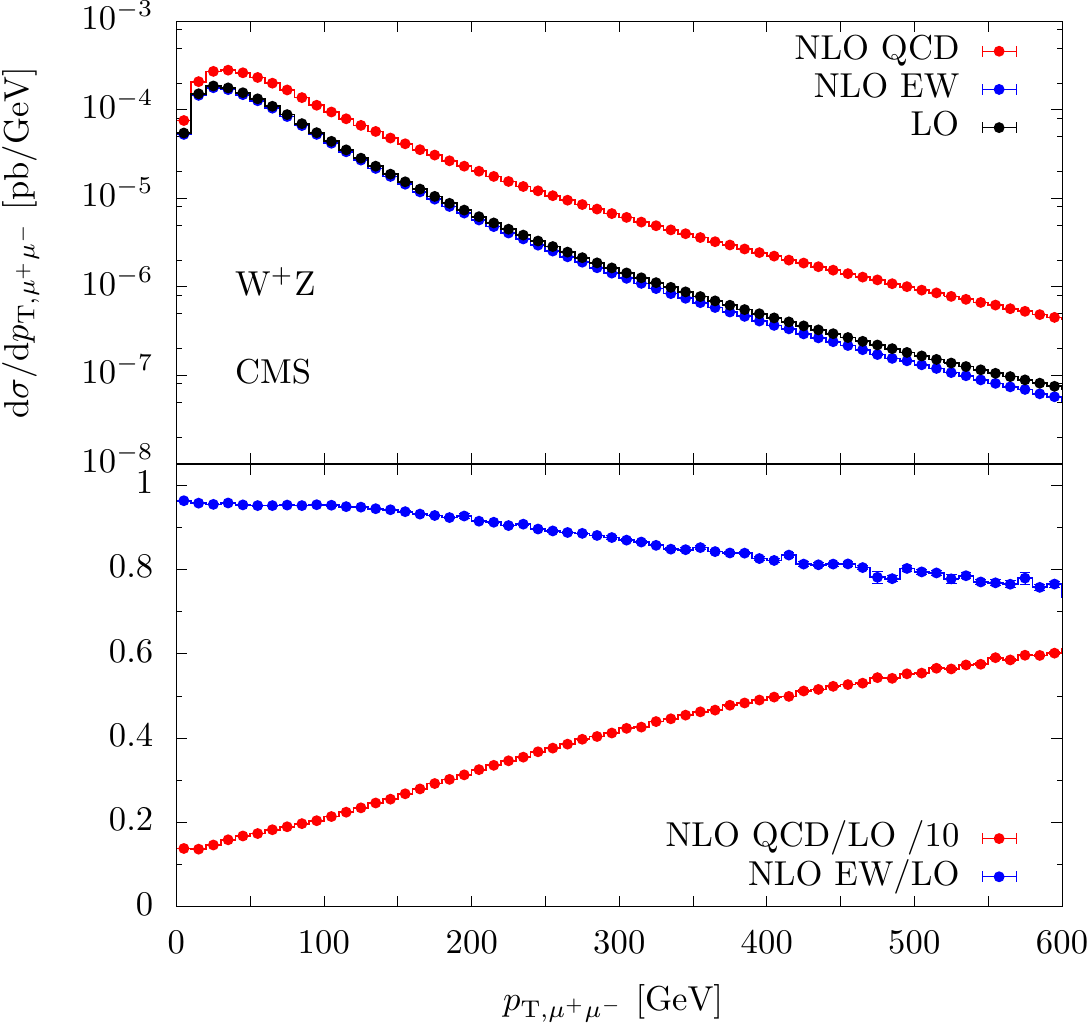}
  \end{minipage}
  \begin{minipage}{0.40\textwidth}
    \includegraphics[width=\textwidth]{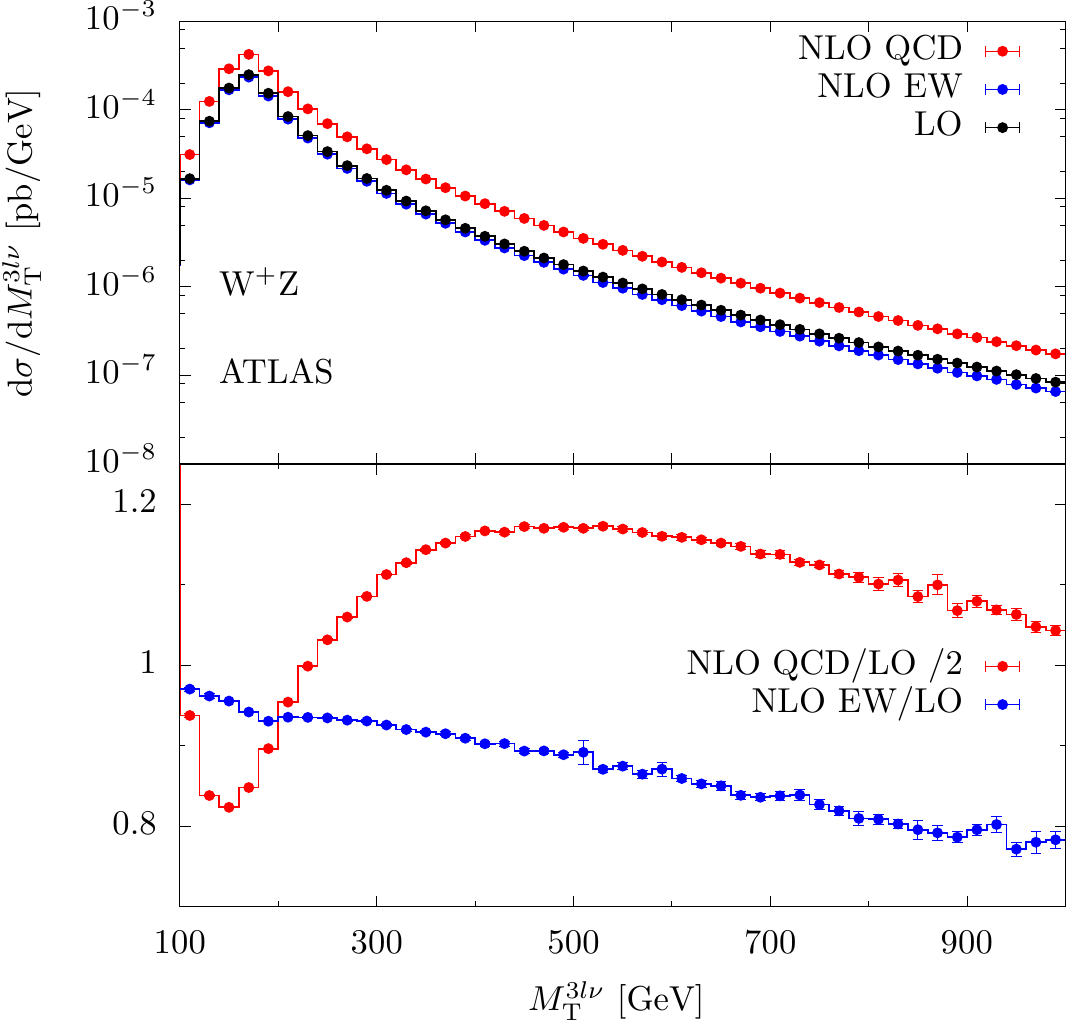}
  \end{minipage}
\end{center}
\caption{Differential distribution in the transverse momentum of the muon--antimuon
  pair ($p_{ {\rm T,}\mu^+ \mu^-}$) and in the $\PW\PZ$ transverse
  mass ($M_{\rm T}^{3l\nu}$) for the process $\Pp\Pp \to \Pep \nu_{\rm e} \mu^+ \mu^-$ at $\sqrt{s}=13$~TeV.
  The NLO QCD   predictions have been divided by a factor 10 (2) in the ratio
  ${\rm    NLO\,QCD}/{\rm LO}$ as a function of $p_{ {\rm T,}\mu^+ \mu^-}$
  ($M_{\rm T}^{3l\nu}$).}
\label{fig:wzptmt}
\efi

Figure~\ref{fig:ratiomt} shows the ratios
\begin{equation}
  \begin{aligned}
  R^{\rm LO (NLO)}_{\rm lin } &=\frac{{\rm d}  \Big( \sigma_{{\rm SM}^2} + \sigma_{{\rm SM}\times {\rm EFT}} \Big)^{{\rm LO(NLO)}\,{\rm QCD}}/{\rm d}X}
  { {\rm d}  \sigma_{{\rm SM}^2}^{{\rm LO(NLO)}\,{\rm QCD}} /{\rm d} X }, \\
  R^{\rm LO (NLO)}_{\rm quad}&=\frac{{\rm d}  \Big( \sigma_{{\rm SM}^2} + \sigma_{{\rm SM}\times {\rm EFT}} + \sigma_{{\rm EFT}^2}\Big)^{{\rm LO(NLO)}\,{\rm QCD}}/{\rm d}X}
  { {\rm d}  \sigma_{{\rm SM}^2}^{{\rm LO(NLO)}\,{\rm QCD}} /{\rm d} X },
  \end{aligned}
  \label{eq:defratio}
\end{equation}
with $X=M_{\rm T}^{3l\nu}$. In the two plots, the upper panels contain only the interference terms
($\sigma_{{\rm SM}\times {\rm EFT6}}$), while in the lower panels also the quadratic terms
($\sigma_{{\rm EFT6}^2}$) are included. Each line corresponds to a setup where only one of
the Wilson coefficients is different from zero. Comparing the predictions for $R^{\rm LO}_{\rm lin}$
and $R^{\rm LO}_{\rm quad}$ reveals that the largest contribution comes from the quadratic terms
(with the only exception of the $c_W$ coefficient, where the linear and the quadratic terms tend
to compensate). Comparing the left and the right plot in Fig.~\ref{fig:ratiomt}, we conclude
that the sensitivity to the Dim-6 operators decreases at NLO QCD (with the exception of the
${\cal O}_{WWW}$ operator, as already noticed in Ref.~\cite{Azatov:2017kzw} for on-shell vector bosons).
The reduction in the sensitivity
is due to the fact that QCD radiation reduces the centre-of-mass energy of the diboson system
with respect to the LO. Since the  contribution of the Dim-6 operators increases with the centre-of-mass
energy of the diboson system, at NLO QCD the contribution of the Dim-6 operators is suppressed.

\bfi
\begin{center}
  \begin{minipage}{0.40\textwidth}
    \includegraphics[width=\textwidth]{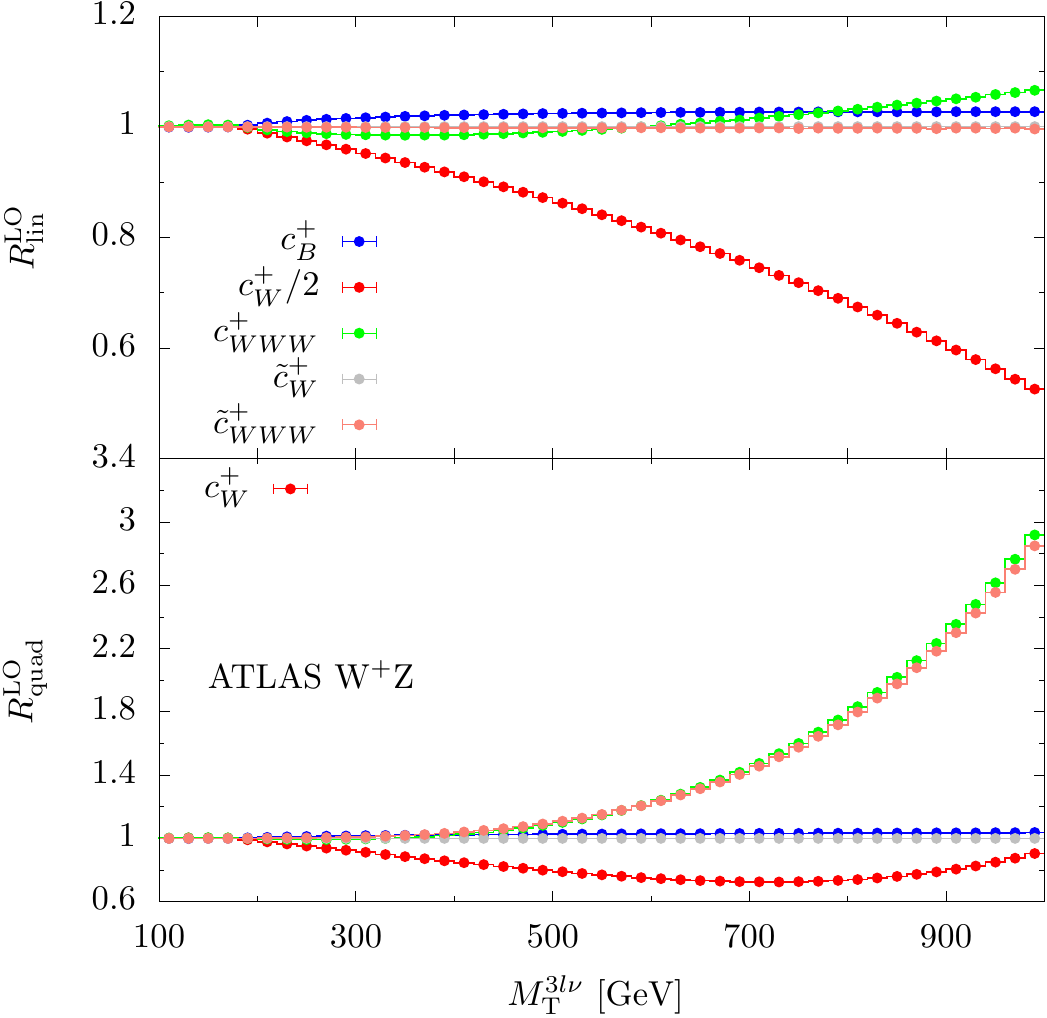}
  \end{minipage}
  \begin{minipage}{0.40\textwidth}
    \includegraphics[width=\textwidth]{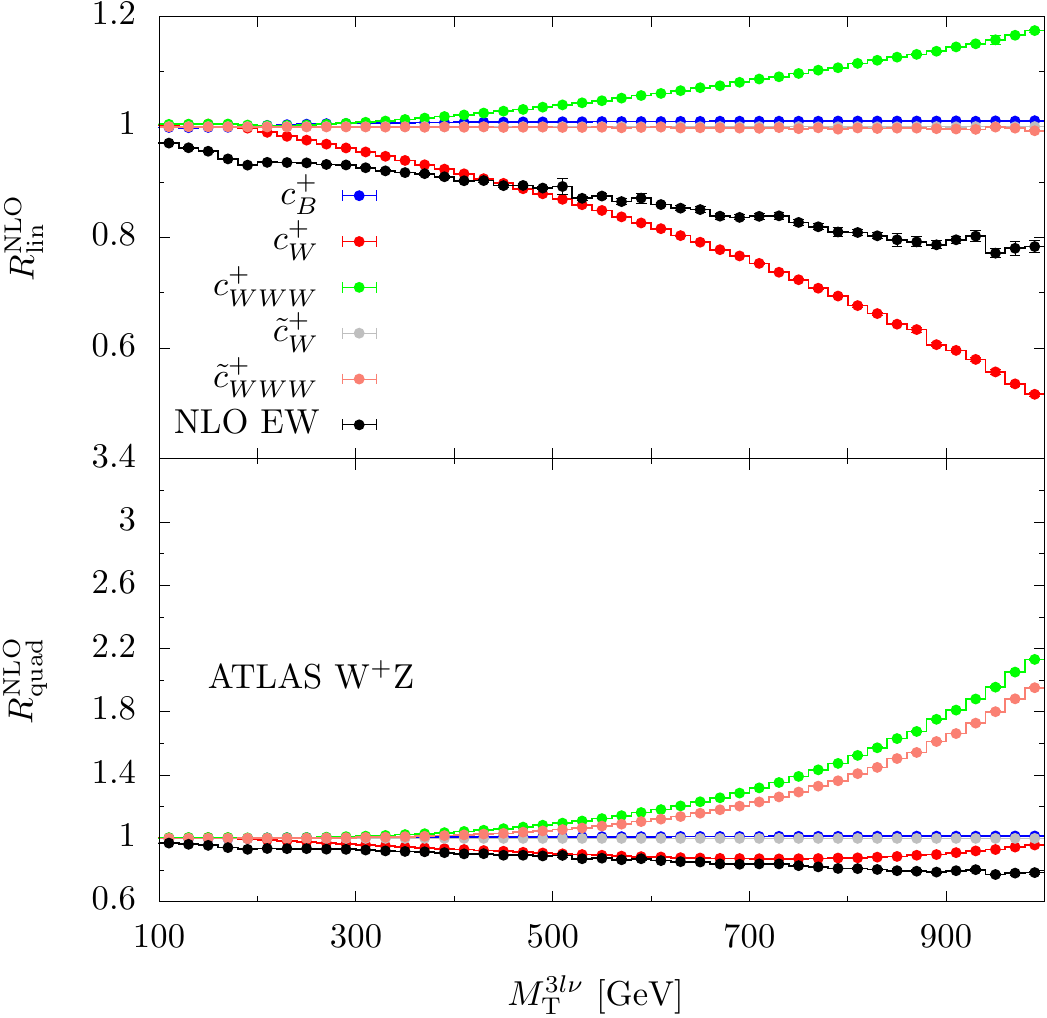}
  \end{minipage}
\end{center}
\caption{Ratio $R^{\rm LO(NLO)}_{\rm lin(quad)}$ as a function of the 
  $\PW\PZ$ transverse mass for the process $\Pp\Pp \to \Pep \nu_{\rm e} \mu^+ \mu^-$. 
  Each line corresponds to a setup where only one of the Wilson coefficients is different
  from zero. The black lines in the right plot correspond to the NLO EW corrections in the SM.
  In order to improve the plot readability, in the $R^{\rm LO}_{\rm lin}$
  ratio (upper panel, left plot) the curve labeled with
  $c_{W}^+/2$ corresponds to our predictions where the
  $c_{W}^+$ coefficient  
  has been divided by a factor 2.}
\label{fig:ratiomt}
\efi

For the process $\Pp\Pp \to  \PZ\PZ \to \Pep \Pem\mu^+ \mu^-$ we consider an event selection that mimics
the ATLAS analysis in Ref.~\cite{Aaboud:2016urj}. Our predictions at the cross-section level are collected in Tab.~\ref{tab:zz-xsec}:
the NLO EW corrections are of order $-8\%$, the NLO QCD corrections are about $+35\%$, and the contribution of the
loop-induced $\Pg\Pg$ channel is of order $+17\%$.

Figure~\ref{fig:zzptm4l} shows the differential distributions as a function of the $p_{\rm T}$ of the hardest Z boson.  
($p_{{\rm T,\,Z}}^{\rm max}$) and of the four-lepton invariant mass ($M_{4l}^{\rm inv}$). The NLO EW
corrections are negative almost everywhere in the plots and reach the value of $-50\%$ and $-45\%$
in the tails of the $p_{{\rm T,\,Z}}^{\rm max}$ and $M_{4l}^{\rm inv}$ distributions, respectively.
The NLO QCD corrections are positive and large, even though not as large as in the case of WZ production.
The $M_{4l}^{\rm inv}$ distribution is less sensitive to the NLO QCD corrections with respect to the
$p_{\rm T}$ of the hardest Z boson.

\begin{table}
  \begin{center}
\renewcommand{\arraystretch}{1.4}
    \begin{tabular}{|l|l|l|l|l}
      \hline
       LO [fb] & NLO QCD [fb] & NLO EW [fb] & $\Pg\Pg$  [fb] \\
      \hline
       $11.0768(5)^{+6.3\%}_{-7.5\%}$ & $14.993(2)^{+3.1\%}_{-2.4\%}$ & $10.283(2)^{+6.4\%}_{-7.6\%}$ & $1.8584(4)^{+25\%}_{-18\%}$\\
      \hline

    \end{tabular}
  \end{center} 
  \caption{Fiducial cross section for the process $\Pp\Pp \to \Pep \Pem \mu^+ \mu^-$
    at $\sqrt{s}=13$~TeV. Same notation and conventions as in Tab.~\ref{tab:wz-xsec}.}
  \label{tab:zz-xsec}
\end{table}

\bfi
\begin{center}
  \begin{minipage}{0.40\textwidth}
    \includegraphics[width=\textwidth]{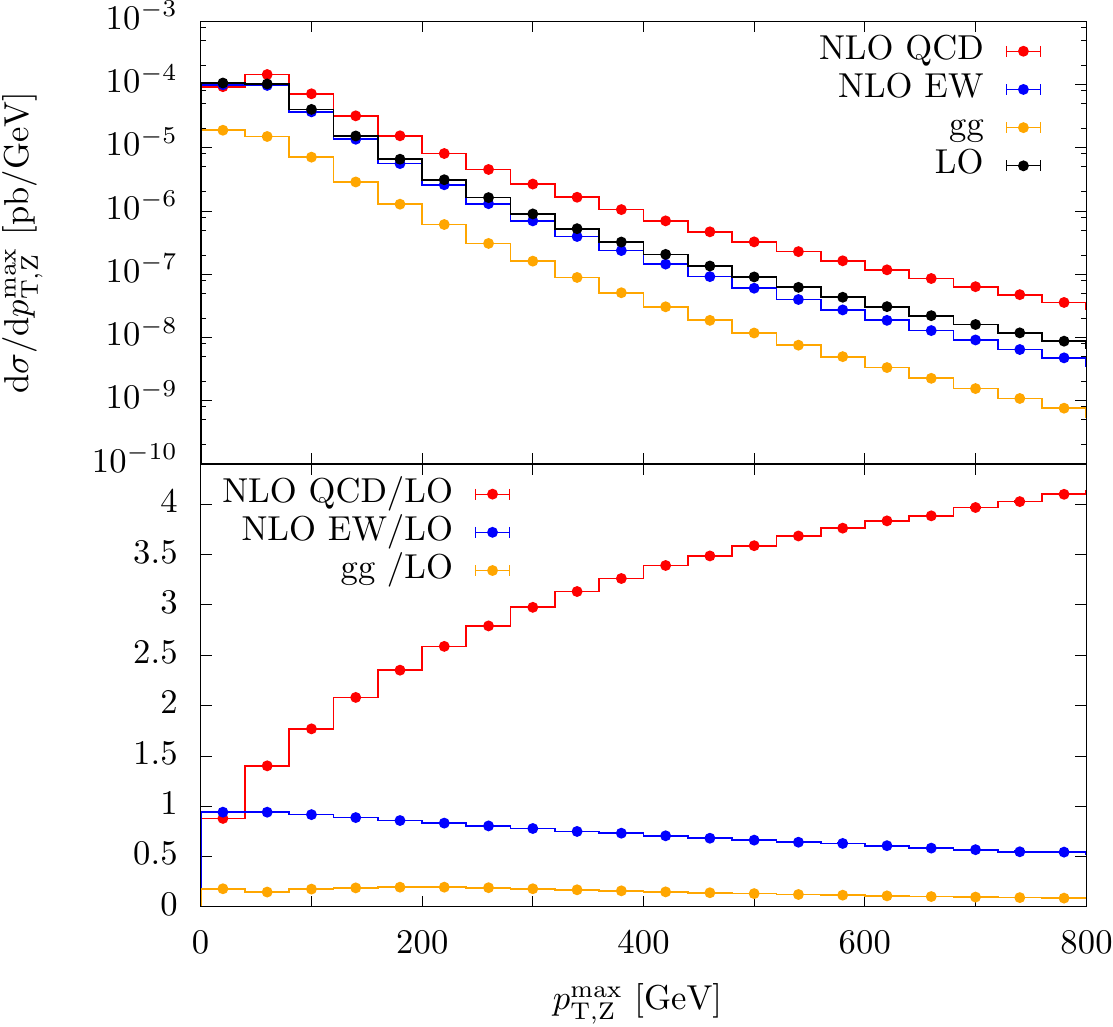}
  \end{minipage}
  \begin{minipage}{0.40\textwidth}
    \includegraphics[width=\textwidth]{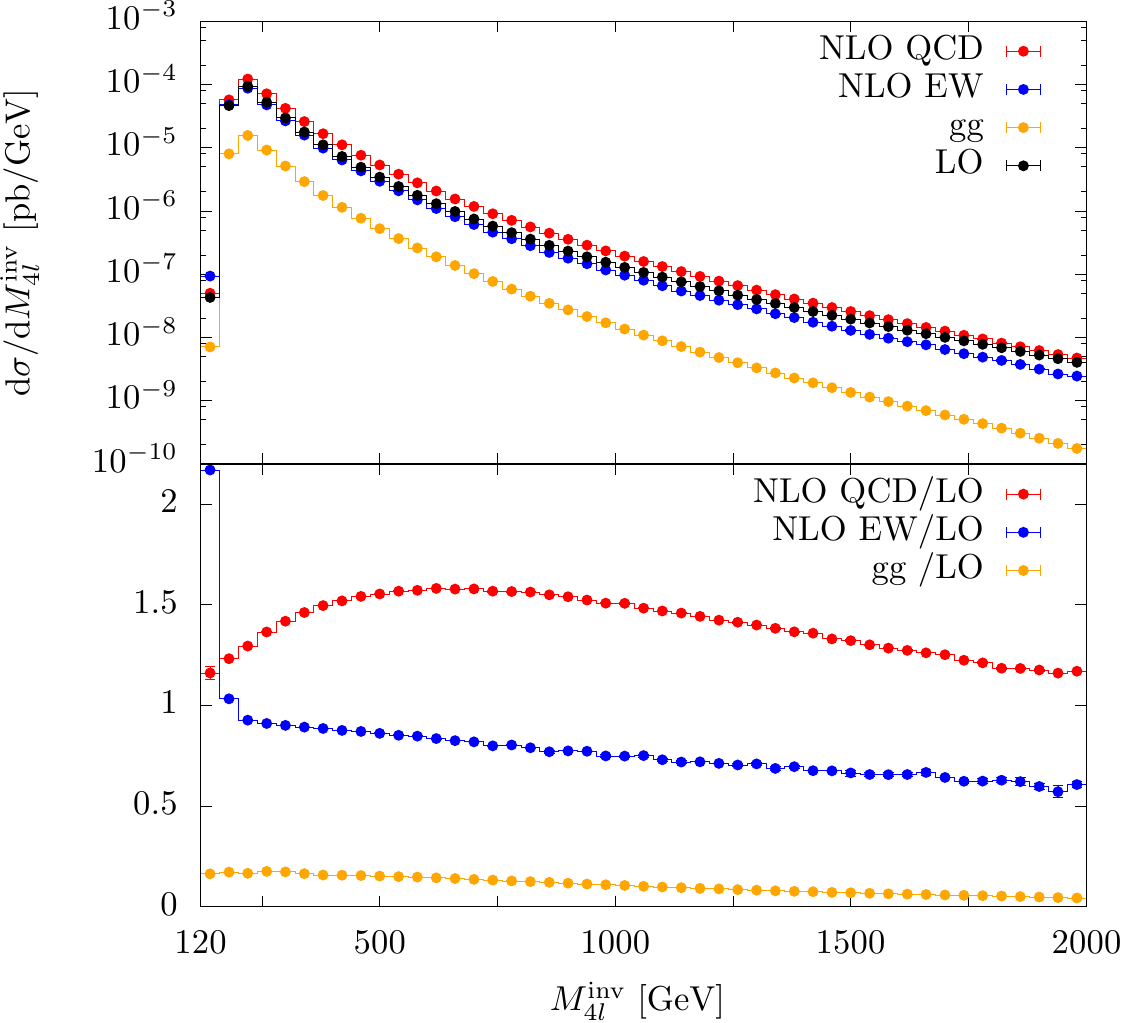}
  \end{minipage}
\end{center}
\caption{Differential distribution in the hardest $\PZ$-boson
  transverse momentum ($p_{{\rm T,\,Z}}^{\rm max}$) and in the four-lepton invariant mass
  ($M_{4l}^{\rm inv}$) for the process
  $\Pp\Pp \to \Pep \Pem \mu^+ \mu^-$  at $\sqrt{s}=13$~TeV.}
\label{fig:zzptm4l}
\efi

\bfi
\begin{center}
  \begin{minipage}{0.40\textwidth}
    \includegraphics[width=\textwidth]{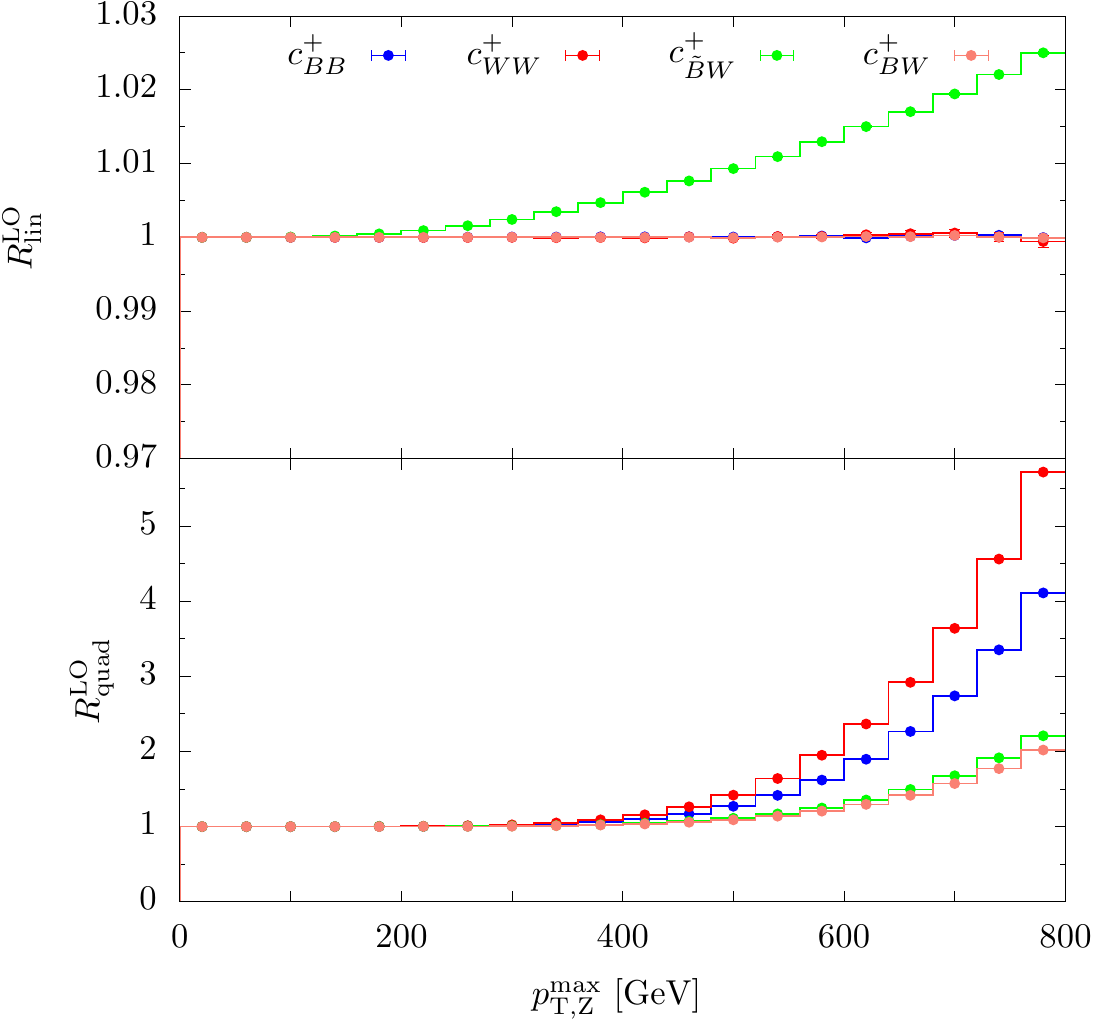}
  \end{minipage}
  \begin{minipage}{0.40\textwidth}
    \includegraphics[width=\textwidth]{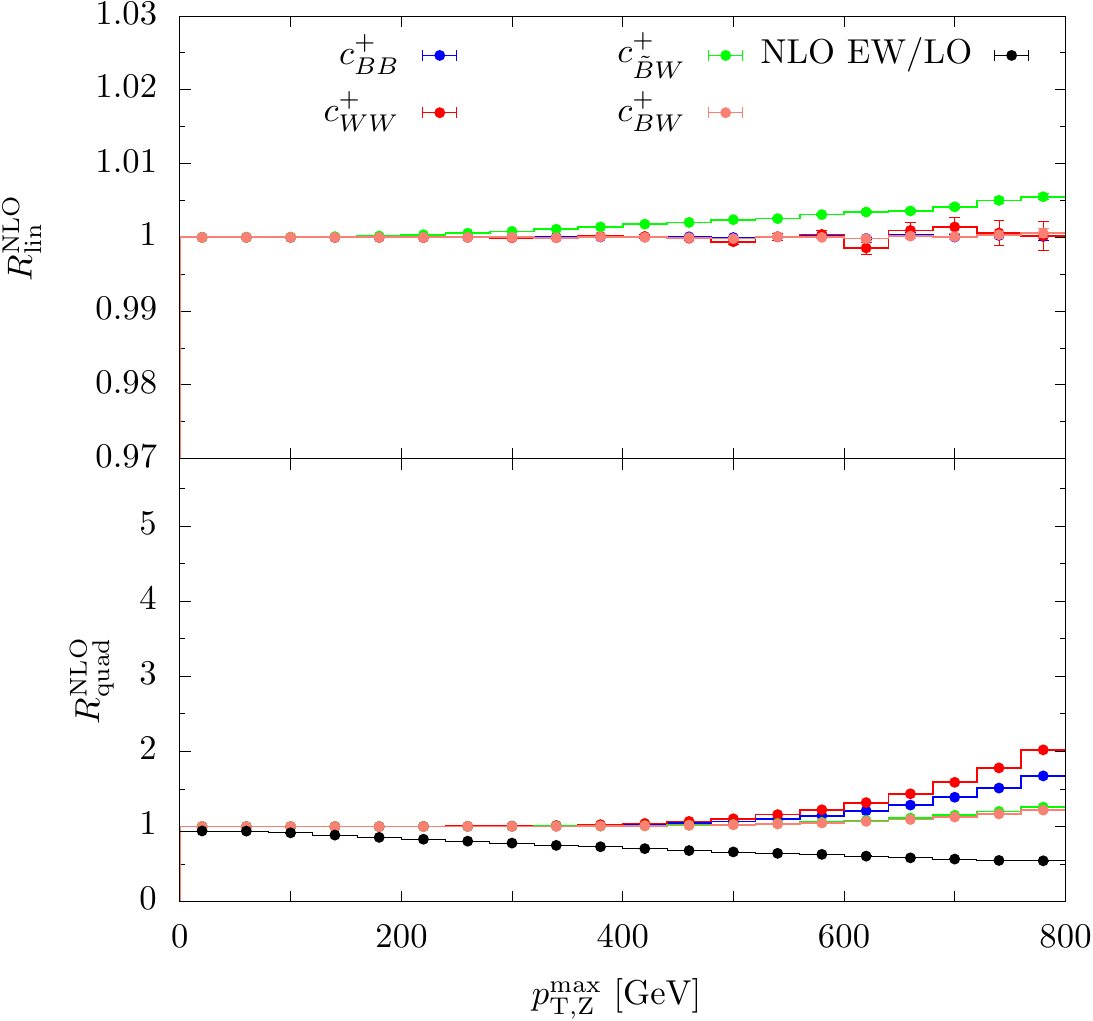}
  \end{minipage}
\end{center}
\caption{Ratio $R^{\rm LO(NLO)}_{\rm lin(quad)}$ as a function of the 
  hardest Z $p_{\rm T}$ for the process $\Pp\Pp \to \Pep \Pem \mu^+ \mu^-$.
  Same notation and conventions as in Fig.~\ref{fig:ratiomt}.}
\label{fig:ratiopt}
\efi

The ratios $R^{\rm LO (NLO)}_{\rm lin (quad) }$ are shown in Fig.~\ref{fig:ratiopt} as a function of the $p_{\rm T}$ of the
hardest Z boson. From the upper panels of Fig.~\ref{fig:ratiopt}, we notice that $R^{\rm LO (NLO)}_{\rm lin }$ is only
sensitive to the $c_{\tilde{B}W}$ coefficient, while $R^{\rm LO (NLO)}_{\rm quad }$ is sensitive to all the four Dim-8
operators of Eq.~(\ref{eq:ntgcop}). By comparing $R^{\rm LO (NLO)}_{\rm lin }$ and $R^{\rm LO (NLO)}_{\rm quad }$ we also
conclude that the largest contribution comes from the quadratic term ($\sigma^{\rm LO (NLO)}_{{\rm EFT8}^2}$). As in the
case of WZ production, the sensitivity to the higher-dimensional operators is reduced at NLO QCD.

\section{Conclusions}
\label{sec:conclusions}

We presented some selected results from Ref.~\cite{Chiesa:2018lcs}, where the processes
 $\Pp\Pp \to \Pep \Pne \Pmum \bar{\nu}_{\mu}$, 
 $\Pp\Pp \to \Pep \nu_{\rm e} \mu^+ \mu^-$,
 and $\Pp\Pp \to \Pep \Pem \mu^+ \mu^-$ 
 have been studied in the EFT framework at NLO QCD accuracy. The impact of the Dim-6 operators
 (Dim-8 operators for ZZ production) has been compared to the NLO QCD and NLO EW corrections
 in the SM.
 
 We found that the sensitivity to the anomalous triple-gauge-boson interaction is in general reduced at NLO QCD 
 because of real radiation contributions, which are less sensitive to the anomalous interaction.

 Our calculation relies on {\sc RECOLA2} for the automated generation and numerical evaluation of the
 tree-level and one-loop amplitudes. This calculation was the first application of {\sc RECOLA2} in the EFT
 framework.

 \subsection*{Acknowledgements}
 The work of M.C. and A.D. was supported by the German Science
Foundation (DFG) under reference number DE 623/5-1.  J.-N. Lang
acknowledges support from the Swiss National Science Foundation (SNF)
under contract BSCGI0-157722.


\end{document}